\RequirePackage{ifpdf}
\ifpdf % We are running pdfTeX in pdf mode
\documentclass[pdftex]{sigma}
\else
\documentclass{sigma}
\fi

\begin{document}

\allowdisplaybreaks

\renewcommand{\PaperNumber}{012}

\FirstPageHeading

\renewcommand{\thefootnote}{$\star$}

\ShortArticleName{Preon Model and Family Replicated $E_6$
Unif\/ication}

\ArticleName{Preon Model and Family Replicated $\boldsymbol{E_6}$
Unif\/ication\footnote{This paper is a contribution to the
Proceedings of the Seventh International Conference ``Symmetry in
Nonlinear Mathematical Physics'' (June 24--30, 2007, Kyiv,
Ukraine). The full collection is available at
\href{http://www.emis.de/journals/SIGMA/symmetry2007.html}{http://www.emis.de/journals/SIGMA/symmetry2007.html}}}

\Author{Chitta Ranjan DAS~$^\dag$ and Larisa V.
LAPERASHVILI~$^\ddag$}

\AuthorNameForHeading{C.R. Das and L.V. Laperashvili}

\Address{$^\dag$~The Institute of Mathematical Sciences, Chennai,
India} \EmailD{\href{mailto:crdas@imsc.res.in}{crdas@imsc.res.in}}
\URLaddressD{\url{http://crdas.crdas.googlepages.com}} %URL address of First Author

% Address of Second Author
\Address{$^\ddag$~The Institute of Theoretical and Experimental
Physics, Moscow,
 Russia}
\EmailD{\href{mailto:laper@itep.ru}{laper@itep.ru}}

\ArticleDates{Received October 02, 2007, in f\/inal form January
24, 2008; Published online February 02, 2008}

\Abstract{Previously we suggested a new preon model of composite
quark-leptons and bosons with the `f\/lipped' $E_6\times
\widetilde{E_6}$ gauge symmetry group. We assumed that preons are
dyons having both hyper-electric $g$ and hyper-magnetic $\tilde g$
charges, and these preons-dyons are conf\/ined by hyper-magnetic
strings which are an ${\rm\bf N}=1$ supersymmetric non-Abelian
f\/lux tubes created by the condensation of spreons near the
Planck scale. In the present paper we show that the existence of
the three types of strings with tensions $T_k=k T_0$ $(k = 1, 2,
3)$ producing three (and only three) generations of composite
quark-leptons, also provides three generations of composite gauge
bosons (`hyper-gluons') and, as a consequence, predicts the family
replicated $[E_6]^3$ unif\/ication at the scale $\sim
10^{17}$~GeV. This group of unif\/ication has the possibility of
breaking to the group of symmetry: $ [SU(3)_C]^3\times
[SU(2)_L]^3\times [U(1)_Y]^3 \times [U(1)_{(B-L)}]^3$ which
undergoes the breakdown to the Standard Model at lower energies.
Some predictive advantages of the family replicated gauge groups
of symmetry are brief\/ly discussed.}


\Classification{81T10; 81T13; 81V22}

\section{Introduction}\label{sec1}

\subsection{Multiple point principle and AntiGUT}\label{sec1.1}

Up to the present time the vast majority of the available
experimental information in high energy physics is essentially
explained by the Standard Model (SM). The gauge symmetry group in
the SM is: \begin{gather}
   SMG = SU(3)_C \times SU(2)_L \times U(1)_Y.
\label{1} \end{gather} All accelerator physics is in agreement
with the SM, except for neutrino oscillations. Presently only this
neutrino physics, together with astrophysics and cosmology, gives
us any phenomenological evidence for going beyond the SM.

The experiment conf\/irms the existence of three generations
(families) of quarks and leptons in the SM. If there exists also
one right-handed neutrino per family, then the SM contains
48~fermion f\/ields and could be described by the enormous global
group $SU(48)\times U(1)$. But Nature chooses only small subgroups
of this global group. The answer is given by simple principles
(see for example~\cite{1}): the resulting theory has to be (i)
free from anomalies, and (ii) free from bare masses. As a result,
we have simple groups (\ref{1}) of the SM. But the largest
semi-simple groups also are possible.

The extension of the SM with the family replicated gauge group
(FRGG): \begin{gather} G = (SMG)^3 = SMG\times SMG \times SMG =
[SU(3)_C]^3\times [SU(2)_L]^3 \times [U(1)_Y]^3  \label{2}
\end{gather} was suggested in~\cite{2,2a} (see also review
\cite{2b}). The appearance of heavy right-handed neutrinos at
$M_{SS}\approx 10^{15}$ GeV was described by the generalized
FRGG-model \cite{3,3a,3b,4}: \begin{gather} G_{\rm ext} =
[SMG\times U_{(B-L)}(1)]^3 = [SU(3)_Y]^3\times [SU(2)_L]^3 \times
[U(1)_Y]^3 \times [U(1)_{(B-L)}]^3. \label{3} \end{gather}

It was assumed that any new physics appears only near the Planck
scale. Such a ``desert scenario'' was accompanied by the Multiple
Point Principle (MPP) suggested in~\cite{5}.

\begin{figure} \centering
\includegraphics[width=55mm]{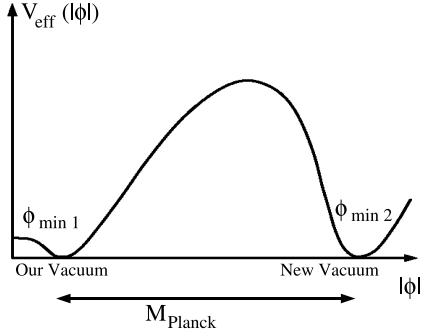}
\caption{The f\/irst (our) vacuum at $|\phi|\approx 246$ GeV and
the second vacuum at the fundamental scale $|\phi|\sim M_{Pl}$.}
\label{f1}
\end{figure}

{\it A priori} it is quite possible for a quantum f\/ield theory
to have several minima of its ef\/fective potential as a function
of its scalar f\/ields. MPP postulates: all the vacua which might
exist in Nature are degenerate and should have approximately zero
vacuum energy density (cosmological constant). According to the
MPP, there are two vacua in the SM (and its extension) with the
same energy density, and all cosmological constants are zero or
approximately zero \cite{6,6a} (see also review \cite{6b}) what is
shown in Fig.~\ref{f1}.

The model \cite{3,3a,3b} in conjunction with MPP \cite{5} f\/its
well the SM fermion masses and mixing angles and describes all
neutrino experimental data \cite{7} (see below
Section~\ref{sec7}). This approach based on the FRGG-model was
previously called Anti-Grand Unif\/ied Theory (AntiGUT).

In the present paper we discuss ``AntiGUT'' as a consequence of
the existence of SUSY GUT~$[E_6]^3$ near the Planck scale, which
is predicted by our new preon model \cite{8,8aa,8a} producing
three generations of composite quark-leptons and bosons. By this
reason, we prefer not to use the name ``AntiGUT" for the FRGG
model~(\ref{3}) and call it simply ``FNT-model''
(Froggatt--Nielsen--Takanishi model \cite{3,3a,3b,7}).

\subsection[Heterotic superstring theory $E_8\times E'_8$]{Heterotic superstring theory $\boldsymbol{E_8\times E'_8}$}\label{sec1.2}

Superstring theory gives the possibility to unify all fundamental
gauge interactions with gravity. The authors of~\cite{9,9a,9b}
have shown that superstrings are free of gravitational and
Yang--Mills anomalies if the superstring theory is described by
the gauge group of symmetry $SO(32)$ or $E_8\times E_8$. A more
realistic candidate for unif\/ication is the ``heterotic''
superstring theory $E_8\times E'_8$ suggested
in~\cite{10,10a,10b}. This ten-dimensional Yang--Mills theory can
undergo a~compactif\/ication. The integration over six
compactif\/ied dimensions leads to the ef\/fective theory with
$E_6$ gauge group of symmetry in four dimensions. The group $E_8$
is broken, but $E'_8$ remains unbroken and plays the role of a
hidden sector in SUGRA. As a result, we obtain the $E_6$ SUSY-GUT
in the four-dimensional space.

\section{A new preon model of composite particles}\label{sec2}

In this paper, as in~\cite{8,8aa,8a}, we present a new model of
preons making composite quark-leptons and (gauge and Higgs) bosons
described by the three types of supersymmetric `f\/lipped' $E_6$
gauge groups of unif\/ication.

We start with the `f\/lipped' supersymmetric group $E_6\times
\widetilde{E_6}$. Here $E_6$ is a non-dual sector of theory with
the hyper-electric charge $g$, and $\widetilde{E_6}$ is a dual
sector with the hyper-magnetic charge~$\tilde g$.

\subsection[Preons are dyons confined by hyper-magnetic strings]{Preons are dyons conf\/ined by hyper-magnetic strings}\label{sec2.1}

The main idea of our investigations, published in~\cite{8,8aa,8a},
is an assumption that preons are dyons conf\/ined by
hyper-magnetic strings which are created by the condensation of
spreons near the Planck scale.

J.~Pati f\/irst suggested \cite{11,11a} to use the strong $U(1)$
magnetic forces to bind preons-dyons in composite objects. This
idea is extended in our model in the light of recent
investigations of composite non-Abelian f\/lux tubes in SQCD
\cite{12,12a,12b,12c,12d}.

Considering the ${\rm\bf N}=1$ supersymmetric f\/lipped $E_6\times
\widetilde {E_6}$ gauge theory for preons in $4D$-dimensional
space-time, we assume that preons $P$ and antipreons $P^{\rm\bf
c}$ are dyons with charges ($ng$, $m\tilde g$) and ($-ng$,
$-m\tilde g$), respectively, residing in the $4D$ hypermultiplets:
\begin{gather*} {\cal H} = (P,P^{\rm\bf c})\end{gather*} and \begin{gather*}  {\cal \tilde H} =
(\tilde P, {\tilde P}^{\rm\bf c}).\end{gather*}  Here ``$\tilde
P$" designates spreons, but not the belonging to
$\widetilde{E_6}$. We assume that the dual sector
$\widetilde{E_6}$ is broken in our world for $\mu \le {\tilde
\mu}_{\rm crit}$ (where $\mu$ is the energy scale) to some group
$\widetilde{G}$.

As a result, near the Planck scale preons and spreons transform
under the hyper-electric gauge group $E_6$ and hyper-magnetic
gauge group $\widetilde{G}$ according to their fundamental
representations: \begin{gather*}
   P,\tilde P\sim (27,N), \qquad  P^{\rm\bf c},\tilde P^{\rm\bf c}\sim
\left(\overline {27},\overline N\right),  % \label{19}
\end{gather*}
where we consider the fundamental representation 27 for $E_6$ and
$N$-plet for $\widetilde{G}$ group. We also take into account
preons and spreons which are singlets of $E_6$: \begin{gather*}
   P_s,\tilde P_s\sim (1,N), \qquad  P_s^{\rm\bf c},\tilde P_s^{\rm\bf c}\sim
\left(1,\overline N\right).
                                        % \label{20}
 \end{gather*}
They are actually necessary for the entire set of composite
quark-leptons and bosons (see~\cite{13}).

The hyper-magnetic interaction is assumed to be responsible for
the formation of~$E_6$ fermions and bosons at the compositeness
scale~$\Lambda_s$.

\subsection[String configurations of composite particles]{String conf\/igurations of composite particles}\label{sec2.2}

Assuming that preons-dyons are conf\/ined by hyper-magnetic
supersymmetric non-Abelian f\/lux tubes which are a generalization
of the well-known Abelian Abrikosov--Nielsen--Olesen (ANO) strings
\cite{14,14a}, we have the following bound states in the limit of
inf\/initely narrow f\/lux tubes (strings):

\begin{itemize}\itemsep=0pt
\item[i)] quark-leptons (fermions belonging to the $E_6$
fundamental representation): \begin{gather*}
            Q^a \sim P^{aA}(y) \left[{\cal P}\exp\left(i\tilde g
\int_x^y\widetilde{A}_{\mu}dx^{\mu}\right)\right]_A^B
                         {({\tilde P}_s^{\rm\bf c})}_B(x) \sim 27,
                                               %\label{21}
\end{gather*} where $ a\in 27$-plet of $E_6$, $A, B\in N$-plet of
$\widetilde{G}$, $\cal P$ is the path ordering and
$\widetilde{A}_{\mu}(x)$ are dual vector potentials belonging to
the adjoint representation of $\widetilde{G}$;

\item[ii)] ``mesons" (hyper-gluons and hyper-Higgses of $E_6$):
\begin{gather*}
            M^a_b \sim P^{aA}(y) \left[{\cal P}\exp\left(i\tilde g
\int_x^y\widetilde{A}_{\mu}dx^{\mu}\right)\right]_A^B
                         (P^{\rm\bf c})_{bB}(x)
 \sim 1+78+650\qquad {\rm of}\  E_6,
                                              % \label{23}
\\
            S \sim({\tilde P}_s)^A(y) \left[{\cal P}\exp\left(i\tilde g
\int_x^y\widetilde{A}_{\mu}dx^{\mu}\right)\right]_A^B
                         {({\tilde P}_{s}^{\rm\bf c})}_B(x) \sim 1;
                                              % \label{24}
\end{gather*}

\item[iii)] for $\widetilde{G}$-triplet we have ``diquarks":
\begin{gather*}
D \sim \epsilon_{ABC}P^{aA'}(z)P^{bB'}(y)({\tilde P}_s)^{C'}(x)
 \left[{\cal P}\exp\left(i\tilde g\int^z_X\widetilde{A}_{\mu}dx^{\mu}
 \right)\right]_{A'}^A \\
 \phantom{D \sim}{}
\times
   \left[{\cal P}\exp\left(i\tilde
g\int^y_X\widetilde{A}_{\mu}dx^{\mu}\right)\right]_{B'}^B
\left[{\cal P}\exp\left(i\tilde
 g\int^x_X\widetilde{A}_{\mu}dx^{\mu}\right)\right]_{C'}^C,
     %\label{26}
     \end{gather*}
and ``baryons": \begin{gather*}
             B \sim
              \epsilon_{ABC}P^{aA'}(z)P^{bB'}(y)P^{cC'}(x)
\left[{\cal P}\exp\left(i\tilde
g\int^z_X\widetilde{A}_{\mu}dx^{\mu}\right)\right]_{A'}^A \\
\phantom{B \sim}{} \times
    \left[{\cal P}\exp\left(i\tilde
g\int^y_X\widetilde{A}_{\mu}dx^{\mu}\right)\right]_{B'}^B
 \left[{\cal P}\exp\left(i\tilde g\int^x_X\widetilde{A}_{\mu}dx^{\mu}
 \right)\right]_{C'}^C.     % \label{27}
\end{gather*}
\end{itemize}

The conjugate composite particles are constructed analogously.

The string conf\/igurations $D$ and $B$ describe a new type of
composite particles belonging to the dif\/ferent representations
of $E_6$.

\begin{figure}[t] \centering
\includegraphics[width=80mm]{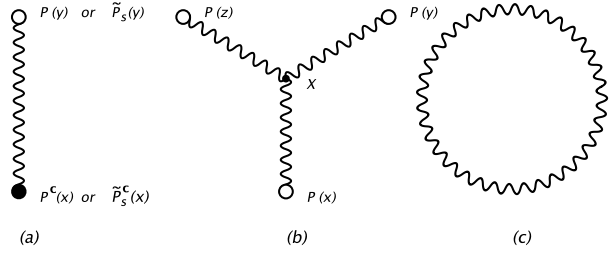}

\caption{Preons are bound by hyper-magnetic strings: (a)
corresponds to the ``unclosed" string conf\/igurations of
composite quark-leptons, hyper-gluons and hyper-Higgses; (b)
corresponds to ``baryo\-nic''  and  ``diquark'' conf\/igurations; (c)
represents a closed string describing a hyper-glueball
(``graviton").} \label{f2}
\end{figure}

The bound states are shown in Fig.~\ref{f2}. It is easy to
generalize these string conf\/igurations for the case of
superpartners~-- squark-sleptons, hyper-gluinos and
hyper-higgsinos. All these bound states belong to $E_6$
representations and they in fact form ${\rm\bf N}=1$ $4D$
superf\/ields.

\section{Condensation of spreons near the Planck scale}\label{sec3}

Let us consider now the breakdown of $E_6$ and $\widetilde{E_6}$
groups at the Planck scale. We assume that Higgses belonging to
the 78-dimensional representation of $E_6$ lead to the following
breakdown: \begin{gather*}
 E_6 \to SU(6)\times SU(2) \to SU(6)\times U(1),
                                                     % \label{28}
\end{gather*} where $SU(6)\times U(1)$ is the largest relevant invariance
group of the 78~\cite{15}. Below we shall show that just this
breakdown provides the spreon condensation and existence of the
second vacuum.

In this investigation, in contrast to our previous
papers~\cite{8,8aa}, we suggest to consider several types of
possible chains for the SM extension by family replicated gauge
groups leading to the $[E_6]^3$ unif\/ication near the scale $\sim
10^{17}$ GeV (according to the predictions of
superstrings~\cite{9}). The chain (explained by the FNT-model
\cite{3,3a,3b,7}): \begin{gather}
 SU(3)_C\times SU(2)_L\times
U(1)_Y \to SU(3)_C\times SU(2)_L\times U(1)_Z \times
 U(1)_X \nonumber\\
 \qquad{}
\to
 [SU(3)_C]^3\times [SU(2)_L]^3\times [U(1)_Z]^3 \times [U(1)_X]^3 \label{28u}
\end{gather}

\begin{itemize}\itemsep=0pt
\item[1)] can be extended by `f\/lipped' models \cite{16}:
 \begin{gather}
 \to [SU(5)\times U(1)_X]^3
 \to  [SU(5)\times U(1)_{Z1} \times
U(1)_{X1}]^3 \nonumber\\
\qquad{} \to [SO(10) \times U(1)_{X1}]^3 \to [E_6]^3, \label{28p}
\end{gather}

\item[2)] or by three ways connected with the left-right symmetry
\cite{17,17a,17b}:
\begin{gather}
 \to [SU(4)_C]^3\times [SU(2)_L]^3 \times [SU(2)_R]^3\times [U(1)_Z]^3\nonumber\\
\qquad{} \to
 [SO(10)\times U(1)_Z]^3 \to [E_6]^3,       \label{28q} \\
  \to [SU(4)_C]^3\times [SU(2)_L]^3 \times [SU(2)_R]^3\times
[U(1)_Z]^3 \nonumber\\
\qquad{}\to [SU(6)\times SU(2)_R]^3 \to [E_6]^3,   \label{28r} \\
 \to [SU(3)_C]^3\times [SU(2)_L]^3 \times [SU(2)_R]^3 \times
[U(1)_X]^3\times [U(1)_Z]^3 \nonumber\\
\qquad{} \to [SU(3)_C]^3\times [SU(3)_L]^3\times [SU(3)_R]^3\to
[E_6]^3.       \label{28s} \end{gather}
\end{itemize}

In the present paper we have investigated only the case 1) of the
`f\/lipped' models. We have chosen this case with the aim to
obtain a minimum of the ef\/fective potential at the scale $\sim
10^{18}$~GeV (see below Section~\ref{sec6}).

We have assumed as an Example~5 of~\cite{16} the breakdown of each
supersymmetric `f\/lipped' $SU(5)\times U(1)_X$ to the
non-supersymmetric gauge group $SU(3)_C\times SU(2)_L\times
U(1)_Z\times U(1)_X$ at the scale $\sim 10^{15}$ GeV, using the
condensates of the Higgs bosons belonging to the $5_h + \bar 5_h$,
$10_H + \overline {10}_H$, $15_{H'} + \overline {15}_{H'}$ and
24-dimensional adjoint A representations of the f\/lipped $SU(5)$.
Then the f\/inal unif\/ication group is the f\/lipped $[E_6]^3$ at
the scale $\sim 10^{17}$ GeV. It is obvious that in this case each
$E_6$ is broken to $SO(10)\times U(1)_{X1}$ by condensates of the
Higgs bosons belonging to the $27_h + \overline {27}_h$, $351_H +
\overline {351}_H$, ${351'}_{H'} + \overline {351'}_{H'}$ and
78-dimensional adjoint A representations of the f\/lipped $E_6$.
In the intermediate region we have the condensates of the Higgs
bosons belonging to the $10_h + \bar 10_h$, $45_H + \overline
{45}_H$, $54_{H'} + \overline {54}_{H'}$ and 45-dimensional
adjoint A representations of the f\/lipped $SO(10)$.

\begin{figure}[t] \centering
\includegraphics[width=90mm,angle=-90]{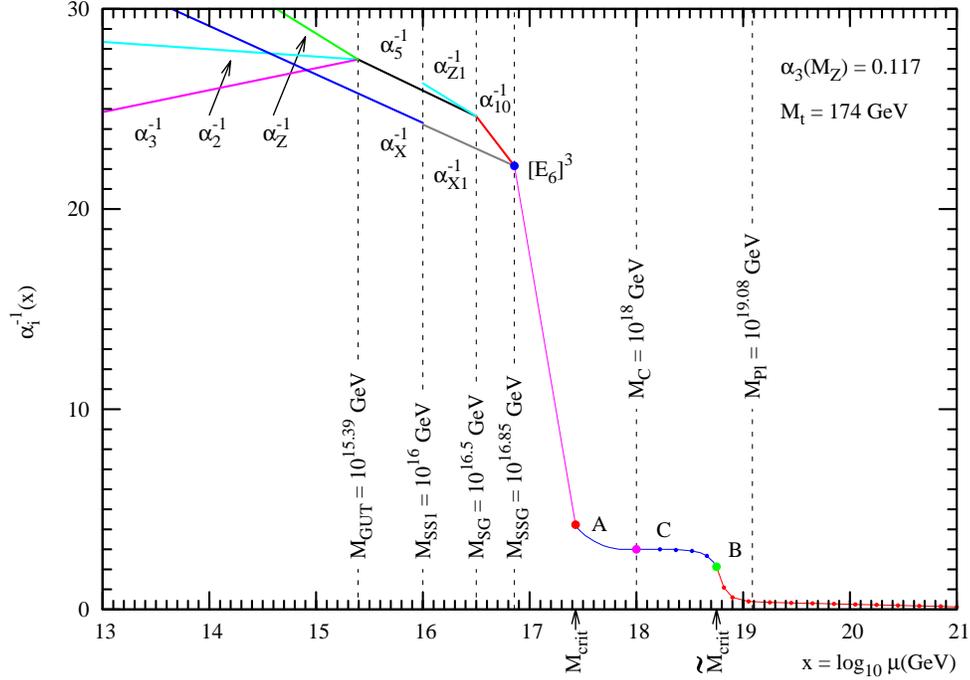}

\caption {The f\/igure provides a qualitative description of the
running of $\alpha^{-1}(x)$ near the Planck scale predicted by the
present preon model. The family replicated gauge group of symmetry
$[E_6]^3$ exists in the region of energies $M_{SSG}\le \mu \le
M_{\rm crit}$.  The point $A$ at the scale of energy $\mu=M_{\rm
crit}$ indicates that hyper-electric preon strings exist for $\mu
\ge M_{\rm crit}$. The point $B$ corresponds to the scale $\mu =
\widetilde{M}_{\rm crit}$ and indicates that hyper-magnetic preon
strings exist for $\mu\le \widetilde{M}_{\rm crit}$. The curve
$AB$ corresponds to the region of spreon condensation, where we
have both hyper-electric and hyper-magnetic strings. The point $C$
corresponds to the second vacuum of our theory. For  $\mu \ge
\widetilde{M}_{\rm crit}$ we have the running of $\alpha^{-1}(x)$
for monopolic ``quark-leptons", but gravity (SUGRA) transforms the
trans-Planckian region. We assume that our theory is valid only up
to the scale $\sim 10^{18}$ GeV (point $C$). } \label{f3}
\end{figure}

Fig.~\ref{f3} presents a qualitative description of the running of
the inverse coupling constants $\alpha_i^{-1}(x)$ near the Planck
scale predicted by the case 1) of our model in the one-loop
approximation of the above-mentioned Example~5 of~\cite{16}. Here
for $\alpha_i=g_i^2/4\pi$ index $i$ corresponds to the
groups~(\ref{28u}) and~(\ref{28p}): $i=1,2,3,X,Z,X1,Z1,5,10$;
$x=\log_{10}\mu$(GeV), and $\mu$ is the energy scale.

Of course, we must understand that the one-loop approximation
running of $\alpha_i^{-1}(x)$ is not valid in the non-perturbative
region $AB$. Two points $A$ and $B$, shown in Fig.~\ref{f3},
correspond (see \cite{8,8aa}): 1)~to the scale $M_{\rm crit}$ of
the breakdown $E_6 \to SU(6)\times U(1)$ (point $A$), and 2)~to
the scale $\widetilde{M}_{\rm crit}$ of the breakdown $\widetilde
{E_6} \to \widetilde {SU(6)}\times \widetilde {U(1)}$ (point~$B$).
We see that near the scale $M_C \approx 10^{18}$ GeV there exists
just the theory of non-Abelian f\/lux tubes, which was developed
recently in~\cite{12,12a,12b,12c,12d}. In contrast
to~\cite{8,8aa,8a}, it is necessary to choose $M_C < M_{Pl}$ with
the aim not to come in conf\/lict with gravity.

We assume the condensation of spreons near the scale $M_C$. One
can combine the $Z_6$ center of~$SU(6)$ with the elements
$\exp(i\pi)\in U(1)$ to get topologically stable string solutions
possessing both windings, in $SU(6)$ and $U(1)$. Henceforth, we
assume the existence of a dual sector of the theory described by
$\widetilde {SU(6)}\times \widetilde {U(1)}$, which is responsible
for hyper-magnetic f\/luxes. Then we have a nontrivial homotopy
group: \begin{gather*}
            \pi_1\left(\frac{SU(6)\times U(1)}{Z_6}\right) = Z_6, %\label{29}
\end{gather*} and f\/lux lines form topologically non-trivial $Z_6$ strings.

Besides $SU(6)$ and $U(1)$ gauge bosons, the model contains thirty
six scalar f\/ields charged with respect to $U(1)$ and $\widetilde
{U(1)}$ which belong to the 6-plets of $SU(6)$ and
$\widetilde{SU(6)}$. Considering scalar f\/ields of spreons
\begin{gather*}
        \tilde P=\left\{ \phi^{aA}\right\},\qquad a,A=1,2,\dots,6, %\label{30}
\end{gather*} we construct their condensation in the vacuum:
\begin{gather*}
       {\tilde P}_{\rm vac} = \big\langle\tilde P^{aA}\big\rangle =
v\cdot {\rm diag}(1,1,\dots,1),\qquad
                               a,A=1,\dots,6.
                               %\label{31}
\end{gather*} The vacuum expectation value (VEV) $v$ is given in~\cite{12} as
\begin{gather*}
              v =\sqrt \xi \gg \Lambda_4,           %\label{32}
\end{gather*} where $\xi$ is the Fayet--Iliopoulos $D$-term parameter in the
${\rm\bf N}=1$ supersymmetric theory and $\Lambda_4$ is its
4-dimensional scale. In our case: \begin{gather*}
             v\sim M_C\sim 10^{18}\,\, {\rm GeV},   %\label{33}
\end{gather*} because spreons are condensed near the scale $M_C$.

Non-trivial topology amounts to the winding of elements of the
matrix \[
        \tilde P=\left\{ \phi^{aA}\right\},\qquad a,A=1,2,\dots,6,\]
and we obtain string solutions of the type:
        \begin{gather*}
 {\tilde P}_{\rm string} = v\cdot{\rm diag}
 \left(e^{i\alpha(x)},e^{i\alpha(x)},\dots,1,1\right),
   \qquad {\rm where}\,\quad x\to \infty. %\label{34}
\end{gather*} Assuming  the existence of a preon $P$ (or spreon $\tilde P$)
and antipreon $P^{\rm\bf c}$ (antispreon $\tilde P^{\rm\bf c}$) at
the ends of strings with hyper-magnetic charges $n\tilde g$ and
$-n\tilde g$, respectively, we obtain the six types of strings
having their f\/luxes $\Phi_n$ quantized according to the $Z_6$
center group of $SU(6)$~\cite{10}: \begin{gather*}
      \Phi_n = n\Phi_0, \qquad n=\pm 1,\pm 2,\pm 3.
                                                      % \label{36}
\end{gather*} The string tensions of these non-Abelian f\/lux tubes were also
calculated. The minimal tension~is: \begin{gather*}
               T_0 = 2\pi \xi,                  %\label{38}
\end{gather*} which in our preon model is equal to: \begin{gather*}
               T_0 = 2\pi v^2\sim 10^{36}\,\, {\rm GeV}^2.
                                                         % \label{39}
\end{gather*} Such an enormously large tension means that preonic strings
have almost inf\/initely small $\alpha' \to 0$, where $\alpha'=
1/(2\pi T_0)$ is the slope of trajectories in the string theory.

The six types of preonic f\/lux tubes oriented in opposite
directions give us the three types of preonic $k$-strings having
the following tensions: \begin{gather*}
   T_k =k\cdot T_0,\qquad {\rm where}\quad k=1,2,3. %\label{39a}
   \end{gather*}
Then hyper-magnetic charges of preons (spreons) and antipreons
(antispreons) are conf\/ined by three types of string.

Also preonic strings are extremely thin. It was shown
in~\cite{8,8aa} that the radius $R_{\rm str}$ of the f\/lux tubes
is very small: $R_{\rm str}\sim 10^{-18}\,\, {\rm GeV}^{-1}$.

\section{Origin of three generations}\label{sec4}

We have obtained three, and only three, generations of fermions
and bosons in the superstring-inspired `f\/lipped' $E_6$ model of
preons. This number ``3" is explained by the existence of just
three values of hyper-magnetic f\/lux tubes which bind the
hyper-magnetic charges of preons-dyons. At the ends of the preonic
strings there are placed hyper-magnetic charges:
 \[\pm \tilde g_0, \qquad {\rm or} \qquad  \pm 2\tilde g_0,\qquad
{\rm or} \qquad \pm 3{\tilde g}_0,\] where ${\tilde g}_0$ is the
minimal hyper-magnetic charge. Then all the bound states form
three generations: for example, three 27-plets of $E_6$
corresponding to the three dif\/ferent tube f\/luxes. We have
obtained a specif\/ic type of ``horizontal symmetry'' explaining
f\/lavor. It was shown  in~\cite{8,8aa} that the model explains
the hierarchy of the SM masses naturally.

 We also have obtained three types of gauge boson $A_{\mu}^i$
(where $i=1,2,3$ is the generation index) belonging to the
$27\times \overline {27} = 1 + 78 + 650$ representations of $E_6$.

\begin{figure}[t]

 \centering
\includegraphics[width=100mm]{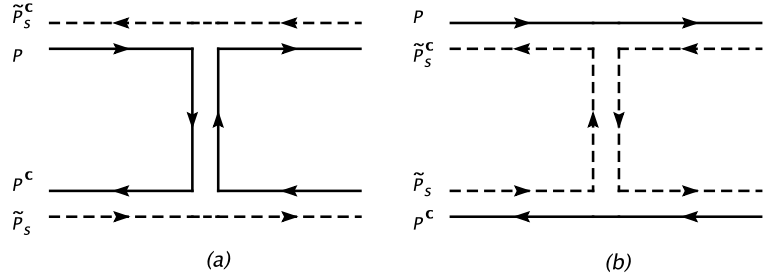}
\caption{Vector gauge bosons belonging to the 78 representation of
the f\/lipped $E_6$ and Higgs scalars~-- singlets of $E_6$~-- are
composite objects created (a) by fermionic preons $P$, $P^{\rm\bf
c}$ and (b) by scalar  spreons ${\tilde P}_s$,~${\tilde P}^{\rm\bf
c}_s$. Both of them are conf\/ined by hyper-magnetic strings.}
\label{f4}
\end{figure}

Fig.~\ref{f4} illustrates the formation of such hyper-gluons
(Fig.~\ref{f4}(a)) and also hyper-Higgses (Fig.~\ref{f4}(b)).

The existence of three generations of hyper-gluons predicts the
family replicated $[E_6]^3$ unif\/i\-ca\-tion near the scale $\sim
10^{17}$ GeV. Here the number of families is equal to the number
of generations: $N_{\rm fam}=N_{\rm gen}=3$. The dynamical
assumption of the three families in our preon model is based on
the existence of the three types of f\/lux tubes (``strings")
connecting preons-dyons in the three (and only three) types of
hyper-gluons which create just the $[E_6]^3$ unif\/ication.

\section[Family replicated ${[E_6]}^3$ unification]{Family replicated $\boldsymbol{[E_6]^3}$ unif\/ication}\label{sec5}

The illustrative picture given in Fig.~\ref{f3} presents the
existence of $[E_6]^3$ unif\/ication in the region of energies
$M_{SSG} \le \mu \le M_{\rm crit}$, where the unif\/ication scale
$M_{SSG}\approx 10^{17}$ GeV.

Here it is necessary to distinguish $E_6$ gauge symmetry group for
preons from $E_6$ for quark-leptons. The points $A$ and $B$ of
Fig.~\ref{f3} respectively correspond to the breakdown of
$[E_6]^3$ and~$[\widetilde{E_6}]^3$ in the region $AB$ of spreon
condensation. In that region we have the breakdown of the preon
(one family) $E_6$ (or $\widetilde{E_6}$):
\[E_6 \to SU(6)\times U(1) \qquad \left({\rm or}\
                  \widetilde{E_6} \to \widetilde {SU(6)}\times
                                        \widetilde {U(1)}\right).
\]

Fig.~\ref{f3} shows that the group $E_6$ is broken in the region
of energies $\mu \ge M_{\rm crit}$ producing hyper-electric
strings between preons. The point $A$ ($B$) indicates the scale
$M_{\rm crit}$ ($\widetilde{M}_{\rm crit}$) corresponding to the
breakdown of $E_6$ ($\widetilde {E_6}$). At the point $B$
hyper-magnetic strings are produced and exist in the region of
energies $\mu \le \widetilde{M}_{\rm crit}$ conf\/ining
hyper-magnetic charges of preons. As a result, in the region  $\mu
\le M_{\rm crit}$ we see quark-leptons with hyper-electric
charges, but in the region $\mu \ge \widetilde{M}_{\rm crit}$
monopolic ``quark-leptons''~-- particles with hyper-magnetic
charges~-- may exist. However, in this region which is close to
$M_{Pl}$, our theory is not correct: gravity (SUGRA) begins to
work, and monopoles are absent in our world.

The dotted curve in Fig.~\ref{f3} describes the running of
$\alpha^{-1}(\mu)$ for monopolic ``quark-leptons'' created by
preons which are bound by supersymmetric hyper-electric
non-Abelian f\/lux tubes. The curve $AB$ corresponds to the region
of spreon condensation, where we have both hyper-electric and
hyper-magnetic strings. The point $C$ corresponds to the second
vacuum of our theory. For $\mu \ge \widetilde{M}_{\rm crit}$ we
have the running of monopole coupling constant. The corresponding
$\alpha^{-1}(x)$ is shown in Fig.~\ref{f3} by the dotted curve.
But our theory is valid only up to the scale $\sim 10^{18}$ GeV
(point $C$ of Fig.~\ref{f3}). Gravity (SUGRA) transforms the
trans-Planckian region, and we do not know anything about the
behavior of theory in the region of dotted curve.

\subsection[The breakdown of ${[E_6]}^3$ to the FNT-model]{The breakdown of $\boldsymbol{[E_6]^3}$ to the FNT-model}\label{sec5.1}

In general, it is quite possible to obtain the FNT-model
considering  the chain (\ref{28u}), (\ref{28p}) of the `f\/lipped'
models at lower energies.

We may assume the breakdown of the supersymmetric `f\/lipped'
$[SU(5)]^3$ to the non-super\-sym\-met\-ric FRGG: \begin{gather}
[SU(3)_C]^3\times [SU(2)_L]^3\times [U(1)_Z]^3\times [U(1)_X]^3
\label{29u}
\end{gather} at the
scale $\sim 10^{15}$ GeV, what was shown in Fig.~\ref{f3}. Then
the f\/inal unif\/ication group is the f\/lipped $[E_6]^3$ at the
scale $M_{SSG}\approx 10^{17}$~GeV.

With the aim to conf\/irm the FNT-model scenario \cite{3,3a,3b,7},
we must assume that the chain~(\ref{28p}) from $[SU(5)]^3$ to
$[E_6]^3$ is realized in the very narrow interval of energies. For
example, in contrast to Fig.~\ref{f3}, we can consider the
breakdown of the supersymmetric `f\/lipped' $[SU(5)]^3$ to the
non-supersymmetric FRGG (\ref{29u}) at the scale $\sim
10^{17}$~GeV, assuming that $M_{SSG}\approx 10^{17.5}$~GeV.
Shortly speaking, the group of unif\/ication $[E_6]^3$ undergoes
an almost direct breakdown to the FRGG group~(\ref{29u}).

As we have mentioned previously, the condensation of spreons near
the scale $M_C$ predicts the existence of a second minimum of the
ef\/fective potential at the scale $\mu\sim 10^{18}$~GeV (see
Fig.~\ref{f1}), according to the Multiple Point
Principle~\cite{5,6,6a,6b}.

\section[Minimum of the effective potential near the Planck scale]{Minimum of the ef\/fective potential near the Planck scale}\label{sec6}

It is not easy to guess how Nature can choose its path from the SM
to the Planck scale. Dif\/ferent paths essentially depend on the
fact whether the intermediate symmetry groups show asymptotically
free or asymptotically unfree (or depressed) behavior of running
gauge couplings. Such a behavior is connected with the number of
representations of the Higgs bosons providing the breaking of the
intermediate symmetry gauge groups down to the SM (what was
considered in Section~\ref{sec3} for~$SU(5)$ and~$SO(10)$ groups
of the chain~(\ref{28p})).

In our preon model the condensation of spreons is possible only if
we have a minimum of the ef\/fective potential near the Planck
scale. Not each of the paths~(\ref{28u})--(\ref{28s}) can give
such a~minimum. The paths (\ref{28q})--(\ref{28s}) of the case 2)
of Section~\ref{sec3} are presumably asymptotically free and do
not give a~minimum of the ef\/fective potential. As it was shown
in~\cite{16}, the $E_6$-unif\/ication can give such a minimum for
the chain~(\ref{28u}),~(\ref{28p}) if there exist symmetry
breaking Higgs bosons belonging to the representations given in
Section~\ref{sec3} by the Example~5 of~\cite{16}. This is a
simplest example, because more complicated cases can be
considered.

In the non-Abelian theory, one usually starts with a gauge f\/ield
$F_{\mu\nu}(x)$ derivable from a~potential $A_{\mu}(x)$:
\begin{gather*} F_{\mu\nu} = \partial_{\nu}A_{\mu}(x) -
\partial_{\mu}A_{\nu}(x) +
i g [ A_{\mu}(x), A_{\nu}(x) ].%\label{30}
\end{gather*} Considering only gauge
groups with the Lie algebra of $SU(N)$, we have: \begin{gather*}
A_{\mu}(x) =
t^jA_{\mu}^j(x),\qquad  j = 1,\dots,N^2 - 1,%\label{31}
\end{gather*} where $t^j$
are the generators of $SU(N)$ group.

\begin{figure}[t] \centering
\includegraphics[width=80mm,angle=-90]{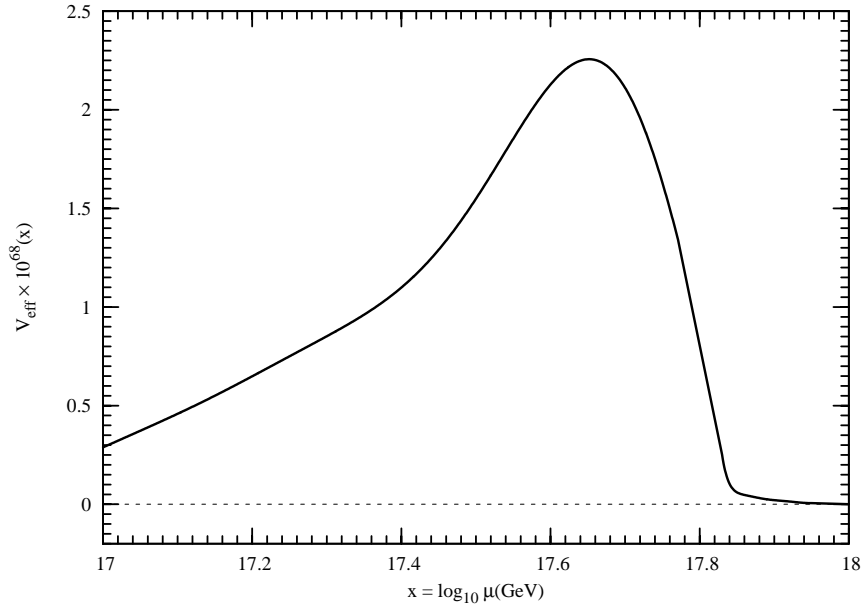}

\caption{The behavior of the ef\/fective potential $V_{\rm
ef\/f}(x)$ for the $E_6^3$ theory, showing a second minimum of our
theory at the point $\mu=M_C=10^{18}$ GeV. This minimum is given
by the requirement: $V_{\rm ef\/f} = V'_{\rm ef\/f} = 0$,
according to the Multiple Point Principle.}\label{f5}
\end{figure}

In general, the perturbative ef\/fective potential is given by the
following expression (see~\cite{18, 19, 19a, 19b}): \begin{gather}
V_{\rm ef\/f}^{(0)} = \frac{\alpha_{\rm
ef\/f}^{-1}(F^2)}{16\pi}F^2 \qquad {\mbox{with}}\quad  F^2\equiv
F_{\mu\nu}^jF^{j\mu\nu} = (\mu\,\, {\mbox{GeV}})^4. \label{32}
\end{gather} Here
$\alpha_{\rm ef\/f}^{-1}(F^2)=\alpha^{-1}(x)$ of our theory.
However, the expression (\ref{32}) is not valid in the
non-perturbative region, because the non-perturbative vacuum
contains a condensation of f\/lux tubes (hyper-electric f\/lux
tubes in our theory of preons), according to the so called
``spaghetti vacuum'' by Nielsen--Olesen~\cite{20}. By this reason,
we subtract the contribution of the condensed f\/luxes from the
expression (\ref{32}): \begin{gather} V_{\rm ef\/f} =
\frac{\alpha_{\rm ef\/f}^{-1}(F^2) - \alpha_{\rm
ef\/f}^{-1}(F_0^2)}{16\pi}F^2,\label{33}
\end{gather} where $F_0^2 =
(M_C\,\, {\mbox{GeV}})^4$. From (\ref{33}) we have:
\begin{gather*}
       V_{\rm ef\/f}(x) = \frac{\alpha^{-1}(x) -
\alpha^{-1}(x_c)}{16\pi}10^{4x}.          % \label{33a}
\end{gather*} The
behavior of the ef\/fective potential $V_{\rm ef\/f}(x)$ is given
in Fig.~\ref{f5}, where we see a second minimum near the Planck
scale at the point $\mu=M_C=10^{18}$~GeV. For this minimum we
have: \begin{gather*} V_{\rm ef\/f}
= V'_{\rm ef\/f} = 0, %\label{34}
\end{gather*} according to the MPP~\cite{5,6,6a,6b}.
But at the scale $M_C$ our theory stops (it is not valid), and we
do not know the development of our theory up to the Planck scale
and further -- in trans-Planckian region.

\section{FNT-model, its advantages and shortcomings}\label{sec7}

If one extends the SM with the FNT-model (\ref{3}) (or with any
FRGG-model) not considering its GUT's origin, then the theory has
some problems. Such a theory

\begin{itemize}\itemsep=0pt
\item[i)] would not account for the quantization of electric (or
magnetic) charge, or for any quantum numbers of the members in a
family, without additional assumptions;

\item[ii)] cannot give a prediction for the weak angle;

\item[iii)] cannot automatically possess B--L as a local symmetry
and the righthanded neutrinos: they are put in by hand.
\end{itemize}

But SUSY GUTs solve these problems in a compelling manner. GUTs
have a predictive power concerning the representations occurring
in the SM. By this reason, it is useful to start with a SUSY GUT
theory explaining the origin of the one or another FRGG-model if
we have indications that this model is valid. SUSY GUT~$[E_6]^3$
predicted by our preon model can be an explanation of the
FNT-model.

However, in a slightly broader way we could think about comparing
of the two competing types of models and see how well they f\/it
and explain the putting of representations for the matter
f\/ields. Such an investigation was published for the FNT-model
in~\cite{3,3a,3b,4,7}. It was shown that 6 dif\/ferent Higgs
f\/ields: $\omega$, $\rho$, $W$, $T$, $\phi_{WS}$, $\phi_{B-L}$
break the FNT-model to the~SM. The f\/ield $\phi_{WS}$ corresponds
to the Weinberg--Salam Higgs f\/ield of Electroweak theory. Its
vacuum expectation value (VEV) is f\/ixed by the Fermi constant:
$\left\langle\phi_{WS}\right\rangle=246$ GeV, so that we have only
5~free parameters~-- f\/ive VEVs:
$\left\langle\omega\right\rangle$,
$\left\langle\rho\right\rangle$, $\left\langle W \right\rangle$,
$\left\langle T \right\rangle$, $\left\langle \phi_{B-L}
\right\rangle$ to f\/it the experiment in the framework of the SM.
These f\/ive adjustable parameters were used with the aim of
f\/inding the best f\/it to experimental data for all fermion
masses and mixing angles in the SM, and also to explain the
neutrino oscillation experiments. It is assumed that the
fundamental Yukawa couplings in our model are of order unity and
so we make order of magnitude predictions.

Experimental results on solar neutrino and atmospheric neutrino
oscillations from Sudbury Neutrino Observatory (SNO Collaboration)
and the Super-Kamiokande Collaboration have been used to extract
the following parameters:
\begin{gather*}
\Delta m^2_{\rm solar} = m_2^2 - m_1^2,\qquad\Delta m^2_{\rm atm}
=
m_3^2 - m_2^2,\\
 \tan^2 \theta_{\rm solar} = \tan^2
\theta_{12},\qquad\tan^2 \theta_{\rm atm} = \tan^2 \theta_{23}
\end{gather*}
where $m_1$, $m_2$, $m_3$ are the hierarchical left-handed
neutrino ef\/fective masses for the three families.

The typical f\/it for the FNT-model is shown in Table~\ref{t1}. As
we can see, the 5 parameter order of magnitude f\/it is
encouraging.

\begin{table}[th]\centering
\caption{Best f\/it to conventional experimental data. All masses
are running masses at $1~\mbox{\rm GeV}$ except the top quark mass
which is the pole mass.}\vspace{1mm}

\begin{tabular}{|c|c|c|}
\hline %\hline
 & {\rm Fitted} & {\rm Experimental} \\ \hline
$m_u$ & 4.4~\mbox{\rm MeV} & 4~\mbox{\rm MeV} \\ \hline $m_d$ &
4.3~\mbox{\rm MeV} & 9~\mbox{\rm MeV} \\ \hline $m_e$ &
1.6~\mbox{\rm MeV} & 0.5~\mbox{\rm MeV} \\ \hline $m_c$ &
0.64~\mbox{\rm GeV} & 1.4~\mbox{\rm GeV} \\ \hline $m_s$ &
295~\mbox{\rm MeV} & 200~\mbox{\rm MeV} \\ \hline $m_{\mu}$ &
111~\mbox{\rm MeV} & 105~\mbox{\rm MeV} \\ \hline $M_t$ &
202~\mbox{\rm GeV} & 180~\mbox{\rm GeV} \\ \hline $m_b$ &
5.7~\mbox{\rm GeV} & 6.3~\mbox{\rm GeV} \\ \hline $m_{\tau}$ &
1.46~\mbox{\rm GeV} & 1.78~\mbox{\rm GeV} \\ \hline $V_{us}$ &
0.11 & 0.22 \\ \hline $V_{cb}$ & 0.026 & 0.041 \\ \hline $V_{ub}$
& 0.0027 & 0.0035 \\ \hline $ \Delta m^2_{\odot}$ & $9.0 \times
10^{-5}~\mbox{\rm eV}^2$ & $ 5.0 \times 10^{-5}~\mbox{\rm eV}^2$ \tsep{0.5ex}\\
\hline $\Delta m^2_{\rm atm}$ & $ 1.7 \times 10^{-3}~\mbox{\rm
eV}^2$ & $2.5 \times 10^{-3}~\mbox{\rm eV}^2$ \tsep{0.5ex}\\
\hline $\tan^2\theta_{\odot}$ &0.26 & 0.34\tsep{0.5ex} \\ \hline
$\tan^2\theta_{\rm atm}$ & 0.65 & 1.0\tsep{0.5ex}\\ \hline
$\tan^2\theta_{\rm chooz}$  & 2.9 $\times 10^{-2}$ & $<2.6 \times 10^{-2}$ \tsep{0.5ex}\\
\hline %\hline
\end{tabular}
\label{t1}

\end{table}

Here attention may drawn to the fact that for quite a long time
now BNP-model (\ref{2}) by Bennett--Nielsen--Picek~\cite{2} and
FNT-model have been to assume that any physics beyond the SM will
f\/irst appear at roughly the Planck scale
(see~\cite{2,2a,2b,5,6b,7,21,22}). A justif\/ication for
continuing to use this so-called ``desert scenario'' could be to
demonstrate that the ef\/fects of the U(1)$_{(B-L)}$ gauge group
associated with the appearance of heavy right-handed neutrinos at
$\sim$ 10$^{15}$ GeV can be neglected for our  study of the values
of the f\/ine structure constants near the Planck scale.

Assuming such a desert, in earlier works invented the MPP/AntiGUT
gauge group model~\cite{2,2a,5} for the purpose of predicting the
Planck scale values of the three Standard Model Group (SMG) gauge
couplings \cite{2,2a,2b,5,6b,7,21,22}, these predictions were made
independently for the three gauge couplings of $SU(3)$, $SU(2)$
and $U(1)$ gauge theories.

According to the BNP-model (\ref{2}), the fundamental group
$SMG^3$ undergoes spontaneous breakdown to the diagonal subgroup
at the energy scale $\mu = \mu_G\sim 10^{17}$ GeV: \begin{gather*}
 SMG^3 \to SMG^3_{\rm diag} = \{g,g,g|g\in SMG\} \cong SMG.  %\label{37}
\end{gather*} For this diagonal subgroup ${(SMG)^3}_{\rm diag}$,
which is identif\/ied with the usual SMG, the gauge couplings are
predicted to coincide with the experimental gauge group couplings
at the Planck scale which in turn are related with critical (i.e.,
multiple point) couplings for $SMG^3$ \cite{5} as follows (see
also~\cite{21,22}):
\begin{gather*}
    \alpha_{1,\exp}^{-1}(\mu_{Pl}) = 6\alpha_{1,{\rm crit}}^{-1},\qquad
  \alpha_{2,\exp}^{-1}(\mu_{Pl}) = 3\alpha_{2,{\rm crit}}^{-1},\qquad
   \alpha_{3,\exp}^{-1}(\mu_{Pl}) = 3\alpha_{3,{\rm crit}}^{-1}.% \label{38}
\end{gather*} Here  $\alpha_i(\mu)$ (where the indices $i=1,2,3$ correspond
in same order to $U(1)$, $SU(2)$ and~$SU(3)$ gauge groups of the
SM) are the SM f\/ine structure constants. Using renormalization
group equations (RGEs) with parameters experimentally established
at the Electroweak (EW) scale, it is possible to extrapolate the
experimental values of the three inverse running
constants~$\alpha_i^{-1}(\mu)$ from EW scale to the Planck scale.
The precision of the LEP data allows us to make this extrapolation
with small errors~\cite{23} even when we ignore the appearance of
the $U(1)_{(B-L)}$ group at the $\mu_{\rm seesaw}\sim 10^{15}$
GeV. Doing the RG extrapolation of the $\alpha^{-1}_i(\mu)$ with
one Higgs doublet and using the assumption of no relevant new
physics up to $\mu \approx \mu_{Pl}$ lead to the following values
(see the Particle Data Group results \cite{23}): \begin{gather}
   \alpha_{1,\exp}^{-1}(\mu_{Pl})\approx 55.4 \pm 6,\qquad
\alpha_{2,\exp}^{-1}(\mu_{Pl})\approx 49.0 \pm 3, \qquad
\alpha_{3,\exp}^{-1}(\mu_{Pl})\approx 53.0 \pm 3. \label{39}
\end{gather} In~\cite{2,2a,2b,5,6b,7,21,22} BNP-model gives an
explanation of these values of the SM coupling constants existing
near the Planck scale.

Also it is necessary to emphasize that FRGG models are extremely
useful for the existence of monopoles in our Universe. Values
(\ref{39}) mean that monopoles are absent in the SM: they have a
huge magnetic charge and are completely conf\/ined or screened.
Supersymmetry does not help to see monopoles.

In theories with the FRGG-symmetry the charge of monopoles $\tilde
g_i$ ($\tilde \alpha = \tilde g^2/{4\pi}$) is essentially
diminished. FRGGs of type $[SU(N)]^{N_{\rm fam}}$ lead to the
lowering of the magnetic charge of the monopole belonging to one
family:
\begin{gather*} \tilde \alpha_{\rm one\,\, family} = \frac{\tilde \alpha}{N_{\rm fam}}.
\end{gather*}  For $N_{\rm fam} = 3$,  for $[SU(2)]^3$ and $[SU(3)]^3$, we
have: \begin{gather*} \tilde \alpha_{\rm one\,\,
family}^{(2,3)}=\frac{{\tilde\alpha}^{(2,3)}}{3}.
\end{gather*} For the family replicated group $[U(1)]^{N_{\rm fam}}$ we obtain:
\begin{gather*} \tilde \alpha_{\rm one\,\, family} = \frac{\tilde \alpha}{N^{*}}
\end{gather*} where $N^{*}=\frac{1}{2}N_{\rm fam}(N_{\rm fam}+1)$. For $N_{\rm fam}=3$
and $[U(1)]^3$, we have: $\tilde \alpha_{\rm one\,\,
family}^{(1)}= {\tilde \alpha}^{(1)}/6$ (six times smaller!). This
result was obtained previously in~\cite{5,24,25}. We conclude:
FRGG models help to observe monopoles in Nature.

Recent investigations \cite{21,22} are devoted to the Planck scale
values of monopole coupling constants. With help of MPP and
FRGG-models we discover new possibilities for monopoles, for dual
charges in general.

\section{Conclusions and discussions}\label{sec8}

In the present paper we have developed a new model of preons
suggested in~\cite{8,8aa,8a}. According to this model, preons
construct composite quark-leptons and (gauge and Higgs) bosons
which are described by the three types of supersymmetric
`f\/lipped' $E_6$ gauge groups of unif\/ication. Crucial points of
this model are: 1) the $E_6$-unif\/ication for preons inspired by
Superstring theory \cite{9,9a,9b,10,10a,10b}, and 2) the dynamical
assumption that preons are dyons conf\/ined by supersymmetric
non-Abelian hyper-magnetic f\/lux tubes of type suggested
in~\cite{12,12a,12b,12c,12d}.

The $E_6$ gauge group of symmetry can be broken into $SU(6)\times
U(2)$ -- from one side, and into $SO(10)\times U(1)$ -- from the
second side. We have assumed that near the Planck scale Higgses
belonging to the 78-dimensional representation of $E_6$ gave the
breakdown $E_6 \to SU(6)\times SU(2) \to SU(6)\times U(1)$, where
$SU(6)\times U(1)$ is the largest relevant invariance group of
the~78~\cite{15}. Assuming the condensation of spreons near the
Planck scale, we have combined the~$Z_6$ center of~$SU(6)$ with
the elements $\exp(i\pi)\in U(1)$ to get topologically stable
string solutions possessing both windings, in $SU(6)$ and $U(1)$.
Just this $Z_6$ center of $SU(6)$ explains the existence of the
three types of preonic $k$-strings associated with the three
generations of quarks-leptons and bosons. Thus, in our preon model
the dynamical properties of preons-dyons predict the three (and
only three) generations (families) of composite quarks-leptons and
bosons. In particular, such a dynamical picture leads to the
existence of the three types of gauge bosons $A_{\mu}^i$ (where
$i=1,2,3$ is the generation index) belonging to the $27\times
\overline {27} = 1 + 78 + 650$ representations of~$E_6$. These
three generations of hyper-gluons predict the family replicated
unif\/ication $[E_6]^3=E_6\times E_6\times E_6$  near the scale
$\sim 10^{17}$ GeV.

In the present paper we have considered the possibility of the
further breakdown of the SUSY GUT $[E_6]^3$ into the
Froggatt--Nielsen--Takanishi (FNT) model $[SU(3)_C]^3\times
[SU(2)_L]^3\times [U(1)_Y]^3 \times [U(1)_{(B-L)}]^3$
\cite{3,3a,3b,7}, existing at lower energies. We have discussed in
Section~\ref{sec7} all shortcomings and advantages of this model.
Among its shortcomings we have no possibility to predict
$U(1)_{B-L}$ gauge group of symmetry and seesaw scale.

We have presented the following advantages of the FNT-model: the
possibility to f\/it the SM parameters (masses and mixing angles),
the description of the Planck scale values of the three SM gauge
couplings, the prediction of the Planck scale values of monopole
couplings etc.

In Section~\ref{sec6} we have presented  investigation of the
existence of the second minimum of the ef\/fective potential in
our preon model, in accordance with the Multiple Point Principle
suggested in~\cite{5,6,6a,6b}.

\subsection*{Acknowledgments}

The authors deeply thank Professor H.B.~Nielsen and one
of the anonymous referees for useful discussions and advices. We are thankful of the
Organizing Committee of the Seventh International Conference
`Symmetry in Nonlinear Mathematical Physics' (Symmetry--2007) (June
24--30, 2007, Kyiv, Ukraine) where our talk was presented.

This work was supported by the Russian Foundation for Basic
Research (RFBR), project No 05--02--17642.

\pdfbookmark[1]{References}{ref}
\LastPageEnding

\end{document}